\documentclass[10pt,twocolumn,a4paper]{esaAI}
\usepackage{subcaption}

\title{Guidance and Control Neural Network Acceleration using Memristors} 

\def\authorEmail{z.a.rudge@tudelft.nl}

\author[1, 2]{Zacharia A. Rudge\thanks{Corresponding author. E-Mail: \authorEmail}}
\author[2]{Dario Izzo}
\author[1]{Moritz Fieback}
\author[1]{Anteneh Gebregiorgis}
\author[1]{Said Hamdioui}
\author[2]{Dominik Dold}
\affil[1]{Delft University of Technology, Delft, the Netherlands}
\affil[2]{ESA Advanced Concepts Team, Noordwijk, the Netherlands}

\begin{document}

\makeCustomtitle

\begin{abstract}

In recent years, the space community has been exploring the possibilities of Artificial Intelligence (AI), specifically Artificial Neural Networks (ANNs), for a variety of on board applications. However, this development is limited by the restricted energy budget of smallsats and cubesats as well as radiation concerns plaguing modern chips. This necessitates research into neural network accelerators capable of meeting these requirements whilst satisfying the compute and performance needs of the application. This paper explores the use of Phase-Change Memory (PCM) and Resistive Random-Access Memory (RRAM) memristors for on-board in-memory computing AI acceleration in space applications. A guidance and control neural network (G\&CNET) accelerated using memristors is simulated in a variety of scenarios and with both device types to evaluate the performance of memristor-based accelerators, considering device non-idealities such as noise and conductance drift. We show that the memristive accelerator is able to learn the expert actions, though challenges remain with the impact of noise on accuracy. We also show that re-training after degradation is able to restore performance to nominal levels. This study provides a foundation for future research into memristor-based AI accelerators for space, highlighting their potential and the need for further investigation.
\end{abstract}
\section{Introduction} \label{sec:intro}
Artificial Intelligence (AI) has the potential to revolutionize space exploration by vastly increasing the autonomy of uncrewed spacecraft, enabling new types of space missions near and far away from Earth alike \cite{furano_towards_2020, izzo_survey_2019,izzo_selected_2023,izzo2022neuromorphic}. Recent advancements have demonstrated the practical application of deep neural networks in various critical tasks, such as real-time optimal control and trajectory generation \cite{cheng_real-time_2020, ma_neural_2025}, autonomous guidance for asteroid landing \cite{ni_accelerating_2023}, illustrating the growing breadth of applications in this domain. However, space applications have strict requirements in terms of accuracy, energy consumption, area, temperature extremes and radiation tolerance \cite{schumann_radiation_2022}. With the introduction of neural networks (NN) to the space domain, and specifically on-board AI accelerators, computational demands continue to grow and these requirements become harder to meet \cite{kothari_final_2020}. 
AI accelerators are typically based on conventional GPU and CPU architectural paradigms, which generally cannot meet these requirements due to, e.g., the \textit{Von Neumann bottleneck} (the energy cost of moving data between the memory and computational units), as well as technology scaling issues and the effect these have on radiation resilience \cite{amrouch_towards_2021}. Moreover, even emerging high-end digital neuromorphic accelerators \cite{schuman_survey_2017} -- such as Intel Loihi \cite{davies_loihi_2018} and FPGA-based implementations -- fail to meet the strict space requirements for similar reasons \cite{montealegre_-flight_2015, schumann_radiation_2022}. Therefore, it becomes of interest to look beyond conventional architectures and device technologies in order to deliver highly energy and area efficient AI solutions that are both sufficiently accurate for the considered space applications and robust to radiation.

Memristor-based in-memory computing is one of the most promising architectures to provide an ideal solution for on-board and edge computing \cite{singh_low-power_2021}. 
It enables the integration of storage and computing within the same location, thus significantly reducing the data-movement \cite{diware_accurate_2023}. 
Not only are memristors non-volatile and exhibit practically no leakage, but they are also resilient to radiation \cite{hughart_comparison_2013} and enable highly parallel computing. This makes the implementation of multiply-accumulate operations (MAC), the core of deep neural networks, extremely efficient \cite{singh_low-power_2021}.
AI accelerators based on memristor device technologies such as PCM (Phase-Change Memory), RRAM (Resistive Random-Access Memory, sometimes also ReRAM) and MRAM (Magnetic Random Access Memory) have been shown to accelerate NN computation while simultaneously reducing energy consumption by orders-of-magnitude when compared to conventional architectures \cite{khaddam-aljameh_hermes_2021, wan_compute--memory_2022, deaville_22nm_2022, bonnet_bringing_2023}. Furthermore, they have been shown to be radiation resistant in a number of scenarios \cite{lyu_research_2021}.
Memristors also create challenges in their implementation, as their usage introduces noise and non-idealities into the computations. Thus far, there has been very limited work done on on-board AI for space applications using emerging technologies such as digital and analog neuromorphic accelerators, \cite{izzo2022neuromorphic,arnold_towards_2024} and -- to the best of our knowledge at the time of writing -- none using memristors. 

This paper advances the state-of-the-art in on-board AI by exploring a new computing paradigm using non-volatile emerging memories. This exploration is done by creating a simulation setup wherein a guidance and control network (G\&CNET) \cite{izzo_stability_2021} is used as a benchmark NN to determine the performance of memristors when used in the context of a space application. This simulation setup is used to experiment with a variety of device parameters and non-idealities (such as conductance drift, read and write noise, bit-slicing and faulty devices) of the simulated hardware and two memristor device technologies, PCM and RRAM \cite{gallo_precision_2022}, with the aim to show the current state of memristor-based accelerators for space applications.

The contributions in this paper are as follows:
\begin{itemize}
    \itemsep0em 
    \item To our knowledge, this is the first work demonstrating a memristor-based NN accelerator for space applications (using PCM / RRAM), in particular a guidance and control NN application. 
    \item We deploy NNs on simulated memristive hardware with different numbers of devices per weight (bit-slicing), from 1 to 16 devices per weight.
    \item We study the impact of device degradation on the accuracy of the NN and the restorative impact of retraining post-degradation.
    \item We further show the impact of conductance drift in the memristors on the accuracy of the NN.
\end{itemize}


The rest of this paper is organized as follows: \cref{sec:methods} describes the experimental setup used in the collection of data for this paper, the neural network by which the hardware is evaluated and why it was chosen, and the details of the experiments performed. \cref{sec:results} shows the results of all experiments performed, and briefly discusses what the results show in terms of performance in various metrics. Finally, \cref{sec:discussion} provides a brief overview of the significance of the results, what other space applications could benefit from memristor-based accelerators, what possible mitigation techniques can be applied to deal with current device non-idealities and avenues for future work.

\section{Methods} \label{sec:methods}
\subsection{G\&CNETs}
To provide an appropriate algorithm for the evaluation of the suitability of a memristor-based neural network accelerator, several target applications were considered. Of these, G\&CNETs were selected as the most suitable NN for this first work on memristor-based computation for space applications.
\begin{figure}[t]
    \centering
    \includegraphics[width=.95\columnwidth]{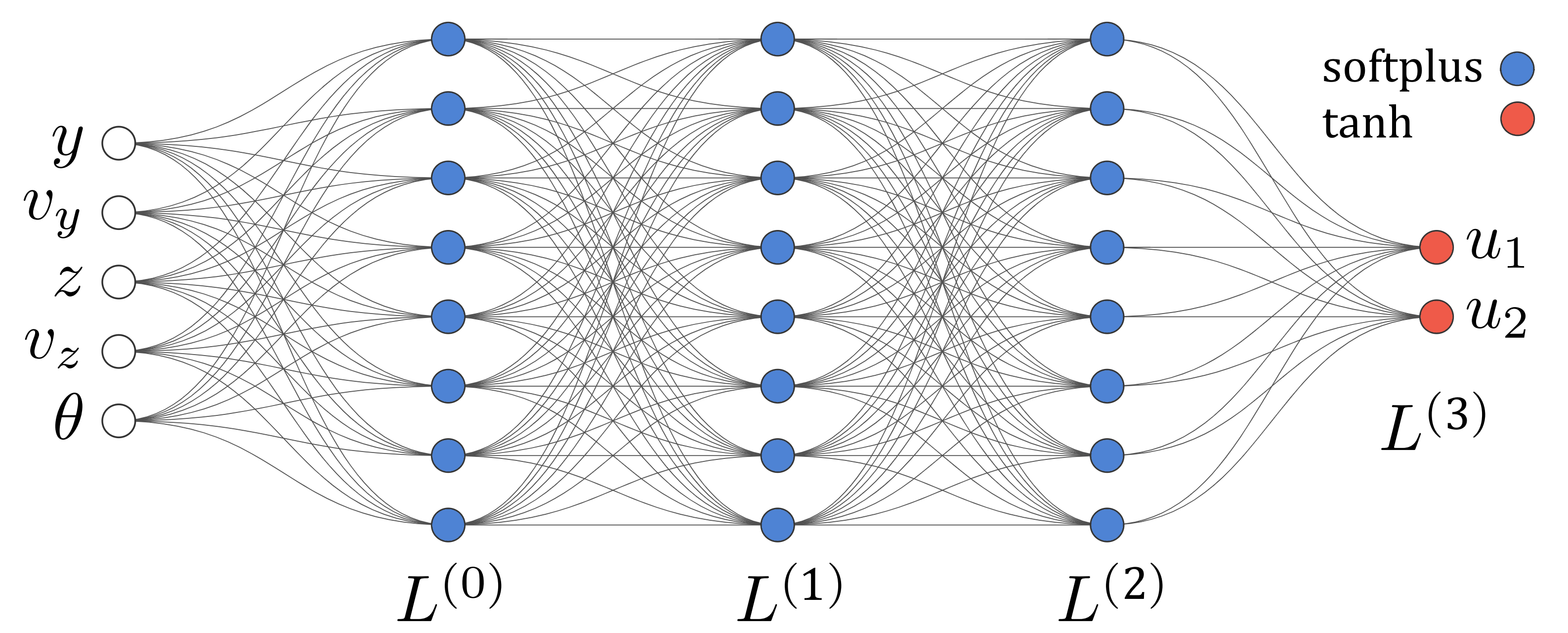}
	\caption{The architecture of the G\&CNET studied. The network is composed of fully connected layers ($L^{(x)}$), input variables (state variables $y$, $v_y$, $z$, $v_z$ and $\theta$) and output variables (optimal control $u_1$, $u_2$).}
	\label{fig:gecnet_arch}\vspace{-5mm}
\end{figure}
G\&CNETs are fully-connected feedforward NNs, often trained using supervised learning on a database of optimal state-action pairs, which are computed by solving from different initial conditions $x_0$, the two-point boundary value problem arising from the application of Pontryagin's maximum principle \cite{pontryagin_mathematical_1987}. 
The resulting network is called a G\&CNET, which is representing the optimal state-feedback: a control $u$, able to steer in time $t_f$ a system from $x_0$ to a target state $x_f$, minimizing the cost function $\mathbf{J}(x,u,t_f) = \int_{t_0}^{t_f}\ell(x,u)dt$. The resulting neurocontrolled system can perform guidance and control in a variety of space applications, thus serving as an on-board substitute for more classical guidance and control systems \cite{izzo_stability_2021}. For the studied NN, we aim to generate and track a trajectory on-board, where the output control vector $U$ reflects the thrust action of a low-thrust spacecraft.

The G\&CNET investigated in the following experiments is a fully-connected feedforward  NN containing three hidden layers with 128 neurons each (\cref{fig:gecnet_arch}). The network takes the state variables of the system as its input, and produces the optimal control $u*$ as its output. The \textit{softplus} and \textit{tanh} activation functions are used in the hidden and output layers, respectively. 
By performing experiments using G\&CNETs, we are able to ascertain the feasibility of using memristor-based AI accelerators for this task, a broad set of other control problems \cite{tailor_learning_2019}, and for other future on-board applications.

\subsection{Memristor-based Neural Network Accelerators}\label{sub:memristor}
To accelerate this network within space application requirements, we look towards memristor-based in-memory computing. Memristors are an emerging technology, the concept of which was first introduced in 1971 \cite{chua_memristor-missing_1971}. It is a two-terminal device which serves as a nonlinear resistor, the resistance $R$ (or inversely the conductance $G$) of which depends on the history of the voltage across it, thus giving way to a ``memory resistor''. The method by which memristors switch their resistance (the so-called ``switching mechanism'') differs per technology used to implement the memristor, but is generally performed by one or more electrical pulses forming and rupturing some form of conductive filament, or changing a magnetic direction \cite{zahoor_resistive_2020}.

\begin{figure}[h]
    \centering
    \includegraphics[trim=0.0cm 0.5cm 0.5cm 0.0cm,width=1.01\columnwidth]{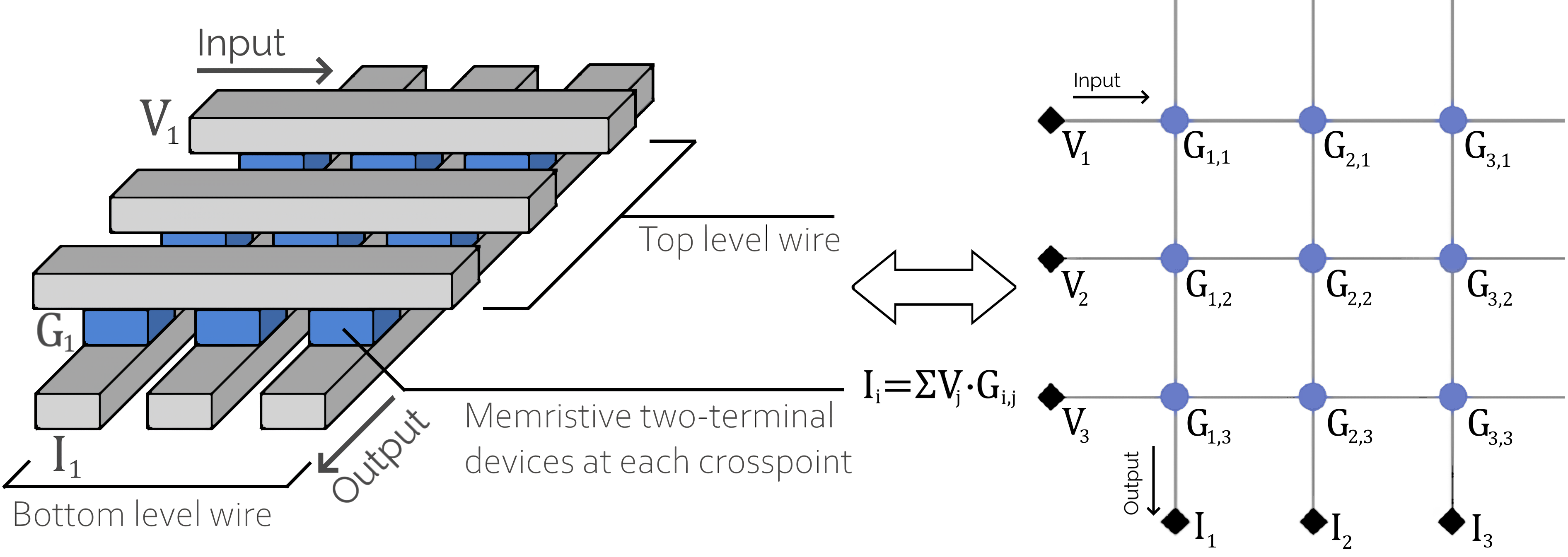}
    \caption{Memristors arranged in crossbar arrays, capable of performing matrix-vector-multiplication (MVM).}
    \label{fig:mca}\vspace{-2.5mm}
\end{figure}

Memristor crossbars (shown in \cref{fig:mca}) can store multiple bits of information in each device, and are able to perform computations with very low energy and latency \cite{ankit_puma_2019}. If an input voltage is applied at the crossbar array's rows, we are able to get the result of a multiplication with the device's resistance as a consequence of Kirchhoff's law, thus performing matrix-vector-multiplication (MVM) in a crossbar array in one computational step. Each element of the crossbar (representing a weight, or element of the vector) is represented by two devices (2R, ``Two Resistors'') in a differential configuration, with one device at 0, and the other device (either the positive or negative) programmed to a value. These memristor crossbar arrays can be built using RRAM \cite{yu_rram_2021}, PCM \cite{wong_phase_2010} and other device technologies. As all calculations are performed in the analog domain, DACs and ADCs are required when interfacing with digital inputs our outputs. Although memristors may be implemented in a radiation resistant manner, it is important to note that the peripheral circuits, which include control blocks and the aforementioned DACs and ADCs, may need radiation hardening when a memristive accelerator is to be used in harsh environments \cite{hughart_comparison_2013, schumann_radiation_2022}. Through the experiments performed we wish to explore various configurations and parameters for a memristor-based NN accelerator. For this we use a simulation setup which allows for modification of values such as the device properties and device technologies as necessary.

\subsection{Simulation Setup}
The memristor-based NN accelerator simulation setup used to perform the experiments is implemented using the \textit{IBM Analog Hardware Acceleration Kit} (IBM AI HW Kit for short) \cite{rasch_flexible_2021}, an open-source Python toolkit that is used to explore in-memory computing using analog devices in the context of NNs in a realistic manner. It is fully integrated within the PyTorch machine learning library for the simulation of training and inference of deep NNs built with PyTorch on analog crossbar arrays. It also features a variety of analog NN modules such as fully connected layers and convolutional layers. Both analog training and hardware-aware training are supported. Furthermore, it is highly customizable in every aspect, ranging from mapping algorithms to supported analog layer types. As part of our contribution to the toolkit, bit-slicing \cite{gallo_precision_2022} (as linear bit sliced layers) has been implemented with the ability to vary the number of slices arbitrarily. 
Bit-slicing is the use of multiple devices to represent the value of one weight in the NN, with each device representing a ``slice'' of a complete weight.

In order to achieve realistic simulation of inference accuracy degradation, the IBM AI HW Kit also provides a specific inference configuration that adds carefully calibrated conductance-dependent programming noise, weight read noise and conductance drift based on a 1M PCM device array. The PCM devices exhibit approximately 2\% read noise on its programmed conductance on average. The same has been done for RRAM devices, based on data from Wan et al.'s work on RRAM-based in-memory computing chips \cite{wan_compute--memory_2022}. The RRAM exhibits 1\% read noise on its programmed conductance on average.

The architecture of the hardware computing the forward pass of the network is as follows: the network consists of multiple layers, each consisting of one or multiple tiles, which contain the memristor crossbar arrays and associated periphery. The weights are mapped to the devices in each tile in a differential configuration, with one device representing a positive synaptic weight and another a negative component, with one of either being set to 0. The peripheral circuitry consists of digital-to-analog converters (DACs, set to a 7-bit resolution) on the input and analog-to-digital converters (ADCs, set to a 9-bit resolution) on the output. It also simulates IR-drop (voltage dropping as current passes through a resistor) and noise from the peripheral circuits (such as from amplifiers used in the ADCs). The result of the computed MVM is then passed to the activation function, which then passes its output to the next layer. 
Lastly, hardware-aware (HWA) training is used to train the memristive NNs. In HWA training, weight updates are calculated in software (i.e., not on-chip) based on inference run on hardware \cite{esser2016,schmitt2017neuromorphic,kungl2019accelerated,rasch_hardware-aware_2023}. In our work, an ideal backward pass is assumed, with the non-idealities only affecting the forward pass. Weight updates derived from this process are then applied to the simulated memristive NN, which is repeated for all training epochs. The code and data to reproduce the results in this paper is available on github \cite{github_repo}.

\vspace{-2.5mm}
\section{Results}\label{sec:results}
To assess the current capabilities of memristive NN accelerators implemented with the device technologies introduced in \cref{sec:intro}, we propose a number of experiments using the simulation setup outlined in \cref{sub:memristor}. The goal of the experiments is to characterize the effect of various properties and non-idealities of these devices on the accuracy of the NN. The accuracy is hereby taken as the average loss of the NN over the validation set. The loss $L$ is defined as the complement of the cosine similarity $S_C$ between the control angles generated by the network and the validation set's optimal angles, in other words: $1-S_c = L$. We compare all results to a baseline of a fully digital floating-point NN. The following experiments were performed: 
\begin{enumerate}
    \itemsep0em 
    \item Introduced slices to the analog tiles and observed the impact of slices on accuracy, ranging from 1 slice per weight to 16 slices per weight.
    \item The accuracy under various levels of device degradation (as a ratio of devices stuck at $G_{min}$, stochastically distributed) was tested, ranging from 1\% to 10\% (with less granular values presented up to 80\%).
    \item The accuracy recovery was observed when re-training is performed post-degradation.
    \item Simulated effect of conductance drift over time on the accuracy of the network (at $t=0s$, $t=1.0s$, $t=24\cdot3600s$, $t=48\cdot3600s$), these values are chosen because realistic data is available from real devices at these values for $t$.
\end{enumerate}

All experiments are performed for both the PCM and RRAM devices, and all experiments aside from the experiment concerning the number of devices per weight (slices) are performed with 8 slices per weight. All experiments were trained for 150 epochs where applicable, including the digital baseline.

\begin{center}
    \begin{figure}[t]
    \begin{subfigure}[t]{.5\linewidth}
      \centering
      \includegraphics[trim=0.22cm 0.22cm 0.1cm 0.22cm, width=0.995\columnwidth]{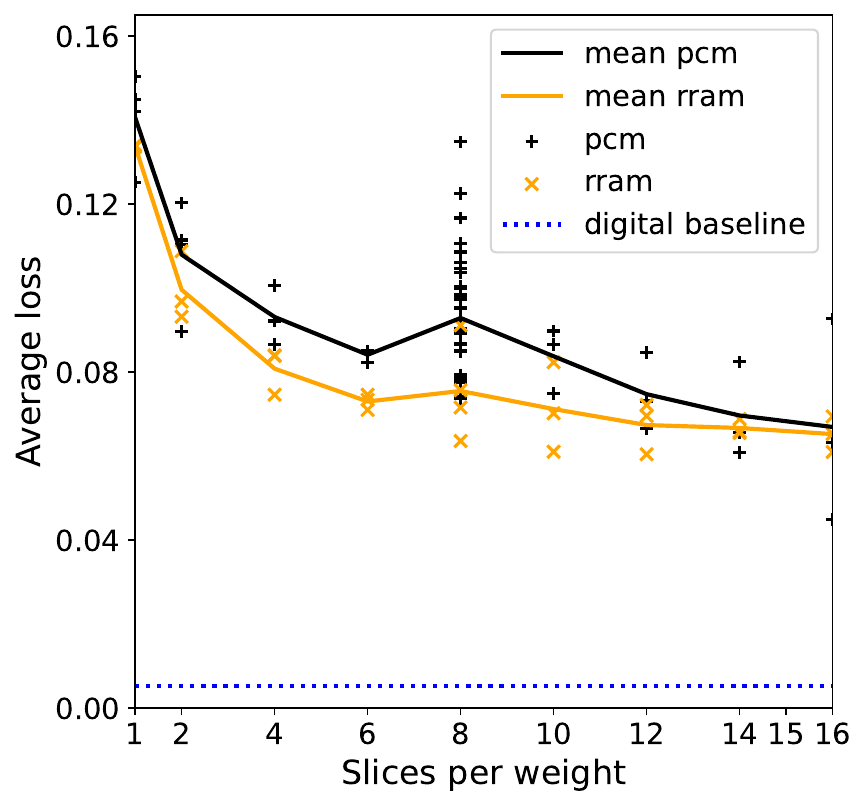}
    \end{subfigure}%
    \hfill
    \begin{subfigure}[t]{.50\linewidth}
      \centering
      \includegraphics[trim=0.22cm 0.22cm 0.1cm 0.22cm,width=0.995\columnwidth]{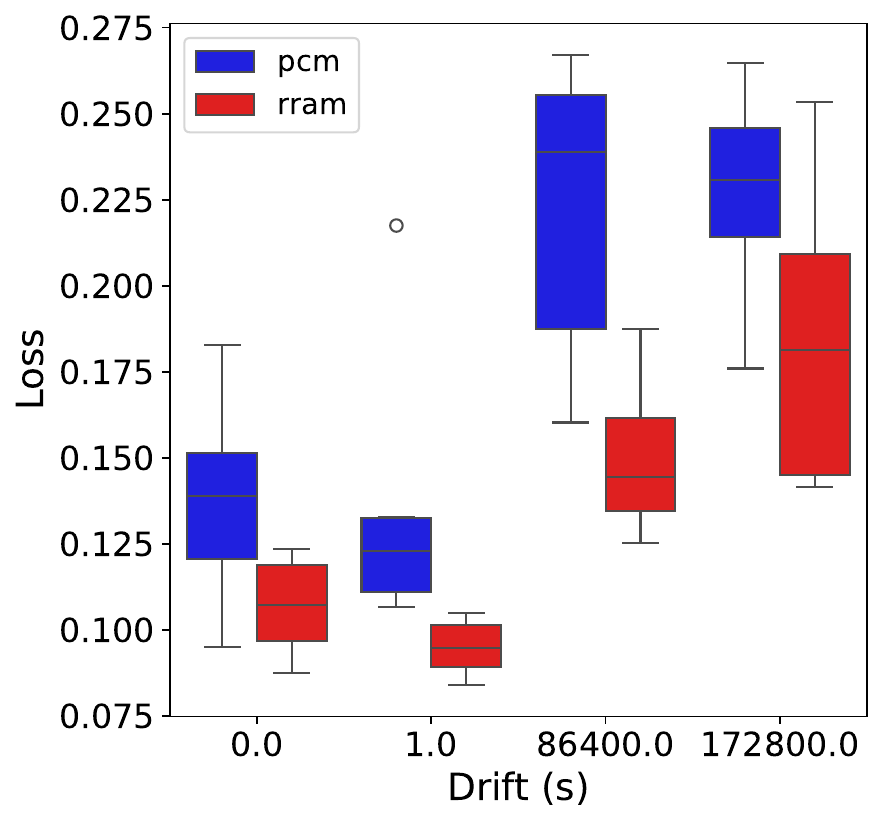}
    \end{subfigure}
    \caption{\textbf{(left)} Loss of model predictions plotted against slices. Digital baseline is shown as a dashed line. \textbf{(right)} The effect of conductance drift on the accuracy of the network.}
    \label{fig:slices_drift}
    \end{figure}\vspace{-7.5mm}
\end{center}

In \cref{fig:slices_drift}~(left), we see that the number of slices has a significant effect on the accuracy of the network, halving the loss (from $\sim0.14$ to $\sim0.07$) when using 8 slices per weight. In increasing the number of slices, we are in effect averaging the noise over multiple devices. We also see diminishing returns in the loss reduction as the number of slices increases towards 16. This is likely due to the fact that as the number of slices increases, the averaging of stochastic noise over multiple devices becomes less effective. For instance, if we assume that the noise of each slice $n_s$ is independent and identically distributed (with finite variance), then the noise amplitude of the weight $n_w$ scales with the number of slices $x$ as $\frac{1}{\sqrt{x}} \cdot n_s = n_w$.
Still, this will have to be studied in more detail in future work.

Secondly, \cref{fig:slices_drift}~(right) shows the effect of conductance drift on the loss of the network. Conductance drift is a non-ideal behavior of memristors, meaning they may lose their programmed value over time even without any applied voltage. ``Refreshing'' the devices to restore the weights is possible, but consumes energy \cite{baroni_tackling_2021}. At time $t=1.0s$, in both the case of the PCM and RRAM devices, the network is mostly unaffected. For PCM, after 24 hours the network's performance has degraded by a factor of $1.4-1.6$. For RRAM, this degradation is more contained, performing approximately $1.3$ times worse. By 48 hours, PCM has doubled in loss, and RRAM has worsened by a factor of $1.5$.


\begin{figure}[b]
    \centering
    \includegraphics[trim=1.2cm 1.2cm 1.2cm 1.2cm, width=0.999\columnwidth]{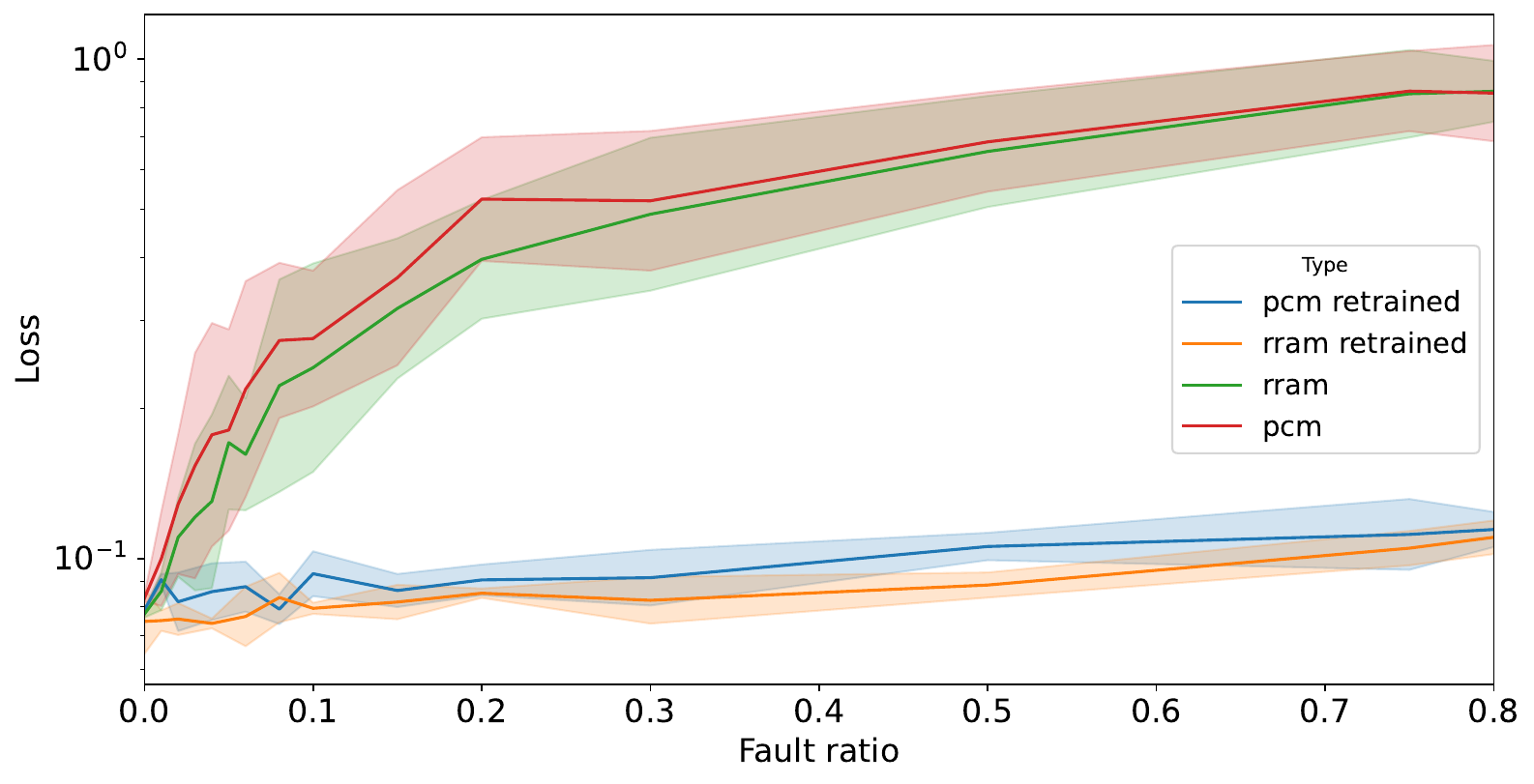}
    \caption{Loss for increasing fault ratios. The two sets of lines depict the PCM and RRAM NNs as affected by faulty devices with (bottom) and without (top) retraining.}
    \label{fig:fault_ratio}\vspace{-2.5mm}
\end{figure}

\Cref{fig:fault_ratio} shows the diminishing performance of the network as the devices fail, with higher fault ratios (above 10\%) showing a loss value typical of a completely untrained network. It also shows the recovery of the degraded memristive NN after re-training, where we can see that performance generally recovers to a large extent (up to a fault ratio of about 10\%). For example, when comparing 10\% device faults without re-training in \cref{fig:fault_ratio} and with re-training, we see that the network recovers significantly from a loss of $\sim0.34$ to a loss of $\sim0.086$.

Finally, the control obtained from the current best analog network (16-slices, RRAM, no faults nor drift, 150 epochs) for a single transfer is shown in \cref{fig:phitheta}, together with the control of the digital baseline NN. More specifically, we show the target spherical coordinates ($\theta$ and $\phi$) as predicted by both the digital and analog NNs -- illustrating that the analog model is capable of performing transfers after training, although with noisier control than the digital model.
%

\begin{center}
    \begin{figure}[h]
    \begin{subfigure}[t]{.5\linewidth}
      \centering
      \includegraphics[clip, trim=9.6cm 0.26cm 7cm .26cm, width=1\columnwidth]{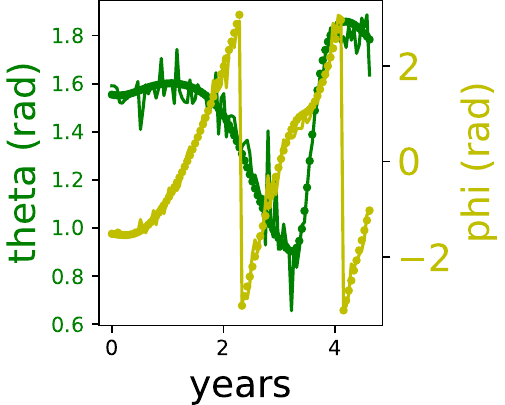}
    \end{subfigure}%
    \hfill
    \begin{subfigure}[t]{.5\linewidth}
      \centering
      \includegraphics[clip, trim=9.6cm 0.26cm 7cm .26cm,width=1\columnwidth]{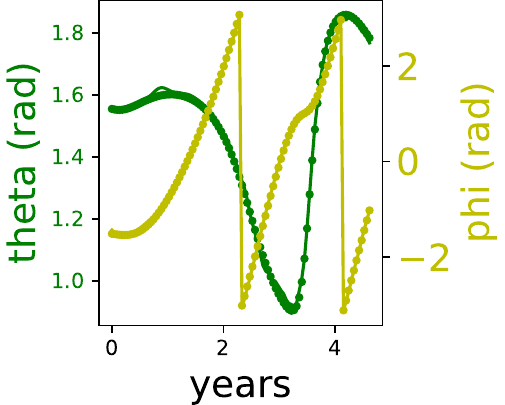}
    \end{subfigure}
    \caption{The network prediction transformed into spherical coordinates $\theta$ and $\phi$ (lines) compared to the optimal ground truth (points). Both the analog (RRAM, left) and digital (right) models are shown.}
    \label{fig:phitheta}
    \end{figure}\vspace{-7.5mm}
\end{center}
%



\section{Discussion}\label{sec:discussion}
We provide -- by way of extensive simulation -- a first glimpse at the current state of memristor-based NN accelerators for space applications.  We found that G\&CNETs, tackling a real-world regression task, can be trained on memristive devices. However, the best loss reached by these models is still  an order of magnitude higher than the one obtained by respective digital models. 
This gap in accuracy seems largely caused by noise during both inference and the programming of weights, since the effects of drift and device faults are either negligible or easily compensated. 
Similarly, we saw that bit slicing has a positive, but also quickly diminishing effect on performance.
Another downside is the noise expressed by the analogue models (\cref{fig:phitheta}) which complicates numerically solving the dynamics of a hypothetical spacecraft controlled by the memristor-based NNs to further evaluate its performance.
Solving these issues and improving accuracy will require extensive research in terms of architectural noise mitigation in hardware, filtering of the output, or improvements in terms of the requirements on the deployed NNs.  

Nevertheless, our results show already now promising performance; even more so within the context of the latency, power and energy reductions enabled by memristors \cite{bonnet_bringing_2023} and their radiation hardness \cite{amrouch_towards_2021}. 
\subsection{Future Work}
There are many avenues in which this research can be expanded, as memristors are a highly promising technology for space applications in particular. One such avenue would be the expansion of the energy estimation capabilities of the simulation setup (for example, by expanding the IBM AI HW Kit) and comparing estimated energy consumption with conventional solutions (such as embedded GPUs and microcontrollers). Furthermore, all networks have been trained for only 150 epochs, for reasons of simulation tractability. 150 epochs is a relatively short training run given that the original G\&CNETs were trained for at least 300 epochs. The shorter training time has affected the accuracy of both the digital and the analog models. A significant problem shown in \cref{sec:results} is the noisiness of the analog network output. One possible method we propose to deal with this is to add a low-pass filter to the output of the NN to smooth the controls. Other ideas include averaging results over ensembles of smaller memristive neural networks, or implementing temporal redundancy by re-running the network (or individual layers) multiple times. 

Lastly, the exploration of other NN space applications and their response to the properties of memristors should be considered. In particular, applications that require on-board learning may be of great interest. One such example would be geodesyNets \cite{izzo_geodesy_2022}, which are networks that learn a three-dimensional, differentiable function representing the density of a target irregular body (such as an asteroid) using minimal prior information. Another example would be adaptive compression algorithms \cite{asiyabi_complex-valued_2023}, which compress SAR (Synthetic Aperture Radar) imagery and perform in-field adaptation to the observed scene. In this situation, memristors would be particularly suitable as they are able to perform low-power inference, and are capable of enabling in-field continuous learning.

This study is purely simulation based, and is thus limited by the data, models and assumptions by which the simulations are performed. Although great effort has been made to ensure realistic assumptions, there will always remain a gap between simulations and reality. We intend to address this in future studies -- for which this paper is a preparatory step -- where hardware will be designed, produced and characterized to evaluate the feasibility of memristive devices for space applications; most of all onboard AI. 


\printbibliography
\addcontentsline{toc}{section}{References}

\end{document}